\documentclass[12pt,preprint]{aastex}

\shorttitle{Seeing double in 30Dor}
\shortauthors{Sabbi et al.}

\begin{document}


\title{A double cluster at the core of 30~Doradus$^1$}

\author{E. Sabbi\altaffilmark{2}, D.J. Lennon\altaffilmark{3}, 
M. Gieles\altaffilmark{4},
S.E. de Mink\altaffilmark{2,5,*},
N.R. Walborn\altaffilmark{2},
J. Anderson\altaffilmark{2},   
A. Bellini\altaffilmark{2}, 
%C.J. Evans\altaffilmark{6,7},  
%V. H\'enault-Brunet\altaffilmark{6}, 
N. Panagia\altaffilmark{2,6,7}, R. van der Marel\altaffilmark{2}, 
J. Ma\'iz Apell\'aniz\altaffilmark{8}}
\email{sabbi@stsci.edu}

\altaffiltext{1}{Based on observations with the NASA/ESA Hubble Space Telescope, obtained at the Space Telescope Science Institute, which is operated by AURA Inc., under NASA contract NAS 5-26555}
\altaffiltext{2}{Space Telescope Science Institute, 3700 San Martin Drive, Baltimore, MD, 21218, USA }
\altaffiltext{3}{ESA/STScI, 3700 San Martin Drive, Baltimore, MD, 21218, USA}
\altaffiltext{4}{Institute of Astronomy, University of Cambridge, Madingley Road, Cambridge, CB3 0HA, UK}
\altaffiltext{5}{Johns Hopkins University, 3400 N. Charles St., Baltimore, MD, 21218, USA}
%\altaffiltext{6}{Scottish Universities Physics Alliance (SUPA), Institute for Astronomy, University of Edinburgh, %Blackford Hill, Edinburgh, EH9 3HJ, UK }
%\altaffiltext{7}{UK Astronomy Technology Center, Royal Observatory Edihburgh, Blackford Hill, Edinburgh, EH9 %3HJ, UK}
\altaffiltext{6}{INAF-CT, Osservatorio Astrofisico di Catania, Via S. Sofia 78, I-95123 Catania, Italy}
\altaffiltext{7}{Supernova Limited, OYV No. 131, Northsound Road, Virgin Gorda, 3700, British Virgin Islands}
\altaffiltext{8}{Instituto de Astrof\'isica de Andaluc\'ia-CSIC, Glorieta de la Astronom\'ia s/n, 18008, Granada, Spain}
\altaffiltext{*}{Hubble Fellow}


\begin{abstract}

Based on an analysis of data obtained with the Wide Field Camera 3 (WFC3) on the Hubble Space Telescope (HST) we report the identification of two distinct stellar populations in the core of the giant H{\sc ii} region 30~Doradus in the Large Magellanic Cloud. The most compact and richest component coincides with the center of R136 and is $\sim 1$ Myr younger than a second more diffuse clump, located $\sim 5.4$ pc toward the northeast. We note that published spectral types of massive stars in these two clumps lend support to the proposed age difference. The morphology and age difference between the two sub-clusters suggests that an ongoing merger may occurring within the core of 30~Doradus. This finding is consistent with the predictions of models of hierarchical fragmentation of turbulent giant molecular clouds, according to which star clusters would be the final products of merging smaller sub-structures.

\end{abstract}


%% subject header.  Objects should be in the appropriate "individual"
%% headers (e.g. quasars: individual, stars: individual, etc.) with the
%% additional provision that the total number of headers, including each
%% individual object, not exceed six.  The \objectname{} macro, and its
%% alias \object{}, is used to mark each object.  The macro takes the object
%% name as its primary argument.  This name will appear in the paper
%% and serve as the link's anchor in the electronic edition if the name
%% is recognized by the data centers.  The macro also takes an optional
%% argument in parentheses in cases where the data center identification
%% differs from what is to be printed in the paper.

\keywords{Magellanic Clouds --- stars: imaging --- galaxies: star clusters: individual --- galaxies: star formation}


\section{Introduction}

As  the most luminous and best known massive star-forming region in the Local Group \citep[LG,][]{Kennicutt1984}, the Tarantula Nebula in the Large Magellanic Cloud (LMC) offers us the unique opportunity to investigate the process of star formation (SF) in an environment that, in many ways (such as metallicity, dust content, and star formation rate), resembles the extreme conditions of the early universe and distant star-forming regions. 
The star formation history (SFH) of Tarantula's ionizing cluster NGC~2070 (a.k.a. 30~Dor) is complex. \citet{Walborn1997} defined 30~Dor as a two-stage starburst, in which the radiation from the compact core R136 has triggered a new generation of stars at a projected distance of $\sim 1\arcmin$ (corresponding to $\sim 15\, {\rm pc}$ at the the distance of the LMC). 

The complexity of NGC~2070 is reflected also in its atypical luminosity profile, which shows a break at  $\sim 10\arcsec$ ($\sim 2.5\, {\rm pc}$) and a bump at $\sim 30\arcsec$ \citep[$\sim 7\, {\rm pc}$;][]{Mackey2003}. More recent observations in the near IR with Multiconjugate Adaptive Optics (MAD) indicate that the innermost $\sim30\arcsec$ of the luminosity profile can be fitted with a single \citet[][EFF]{Elson1987} powerlaw, once the asymmetric shape of 30~Dor is taken into account \citep{Campbell2010}. 

Here we take advantage of the high spatial resolution and sensitivity of the Wide Field Camera 3 (WFC3) on board of the {\it Hubble} Space Telescope (HST) to study the innermost $\sim2\arcmin 44\arcsec\times 2\arcmin 44\arcsec$ of NGC~2070. In particular, we use public UVIS and IR channel observations (P.I. R.W. O'Connell, GO-11360) to characterize the spatial distribution of its stellar populations.

\section{Data}

Deep WFC3 observations of R136, the core of NGC~2070, were retrieved from the Mikulski Archive for Space Telescopes (MAST). The data-set consists of 6 broadband filters (F336W, F438W, F555W, F814W, F110W, and F160W), ranging from the near-UV to the near-IR, and is described in detail in \citet{deMarchi2011}. The images were processed through the standard calibration pipeline CALWF3, using the most up-to-date calibration files. 

Positions and fluxes of the stars in the WFC3/UVIS images were measured using img2xym\_wfc3uv, while we used img2xym\_wfc3ir to measure positions and fluxes in the WFC3/IR images. Both codes were adapted from img2xym\_WFI \citep{Anderson2006}. The astrometry and photometry of WFC3/UVIS data were corrected for pixel area variations and geometric distortion as in \citet{Bellini2009} and \citet{Bellini2011}, while for the WFC3/IR data we used the solutions from Anderson et al. (in preparation). Stars were detected independently in each filter. The WFC3 photometric catalogs were calibrated into the Vegamag photometric system using the most up-to-date values of the zeropoints, available at the WFC3 webpage\footnote{http://www.stsci.edu/hst/wfc3/phot\_zp\_lbn}, and then combined into a single catalog. Approximately 3000 stars with photometric errors smaller than $<0.15$ mag were found in all six bands.  

\section{Analysis}

\begin{figure}
\center
\epsscale{1.0}
\plotone{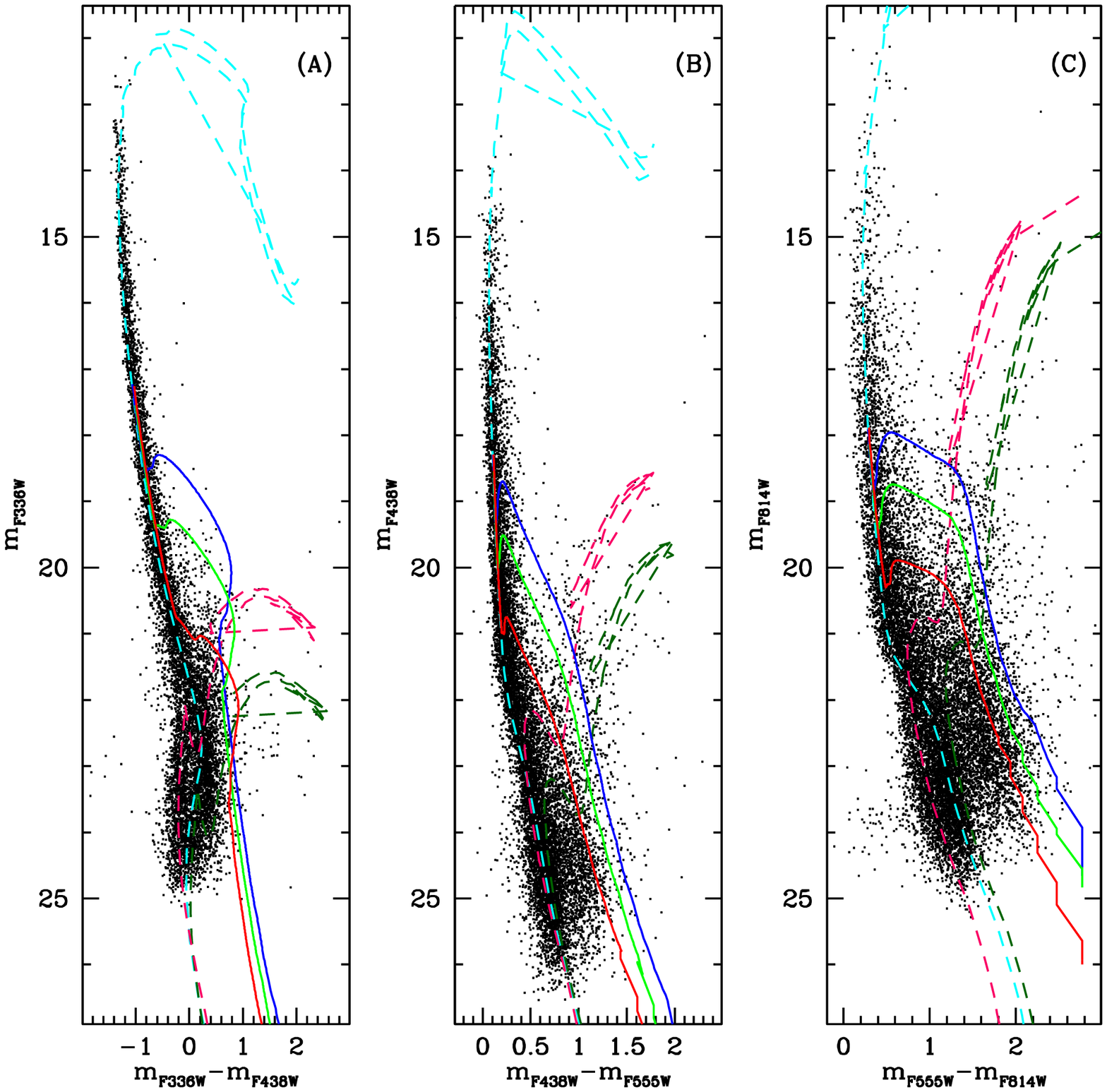}
\caption{\label{3cmds} CMDs of $m_{\rm F336W}$ vs. $m_{\rm F336W}-m_{\rm F438W}$ (panel A),  $m_{\rm F438W}$ vs. $m_{\rm F438W}-m_{\rm F555W}$ (panel B), and $m_{\rm F814W}$ vs. $m_{\rm F555W}-m_{\rm F814W}$ (panel C), of the stars found in the NGC~2070 region with photometric errors $<0.15$ mag in both filters. Pisa isochrones \citep[solid;][]{Tognelli2011} for PMS stars of metallicity Z=0.008 and ages=1, 2 and 5 Myr are shown in blue, green and red respectively.  The dashed isochrones are from \citet{Marigo2008}.  A 10 Myr old stellar population with Z=0.008 is shown in cyan. A 4 Gyr old stellar population with Z$=0.004$ and two distinct values of reddening (E(B-V)=0.14 and 0.4) is shown in magenta and dark green respectively. }
\end{figure}

Fig.~\ref{3cmds} shows three color-magnitude diagrams (CMDs) of the stars found in the region covered by the UVIS observations for different combinations of the filters F336W, F438W, F555W, and F814W. In each plot, stars were selected to have photometric errors $<0.15$ in both the plotted filters. Padova isochrones \citep[dashed lines; ][]{Marigo2008} for main sequence (MS) and evolved stars, and Pisa isochrones \citep[continuous lines; ][]{Tognelli2011} for pre-main sequence (PMS) stars, are superimposed to guide the eye.  In doing this we assumed a distance of 50 kpc \citep{Schaefer2008}, and a  metallicity Z=0.008 \citep{Helling2000}. Isochrones are also corrected for reddening.

Inspection of these CMDs shows that the stellar content is complex: the bright and blue upper-MS (i.e. $m_{\rm F814}\le 21$; $m_{\rm F555W}-m_{\rm F814}\la 0.6$ in panel C) is composed of intermediate and high mass stars, but  at redder colors (i.e. $m_{\rm F555W}-m_{F814W}\ga 1.5$ in panel C) the majority of the sources are PMS stars in the mass range $0.5\la M\la 5.0\, {\rm M_\odot}$ and ages $\ga 1\, {\rm Myr}$.  Pisa isochrones show that PMS stars become fainter and bluer with age, to the point that after $\sim10\,  {\rm Myr}$ they are indistinguishable from the low-mass MS ($m_{\rm F814W}\ge 21$; $m_{\rm F555W}-m_{\rm F814}\la 1.0$ in panel C).

\begin{figure}
\epsscale{0.8}
\plotone{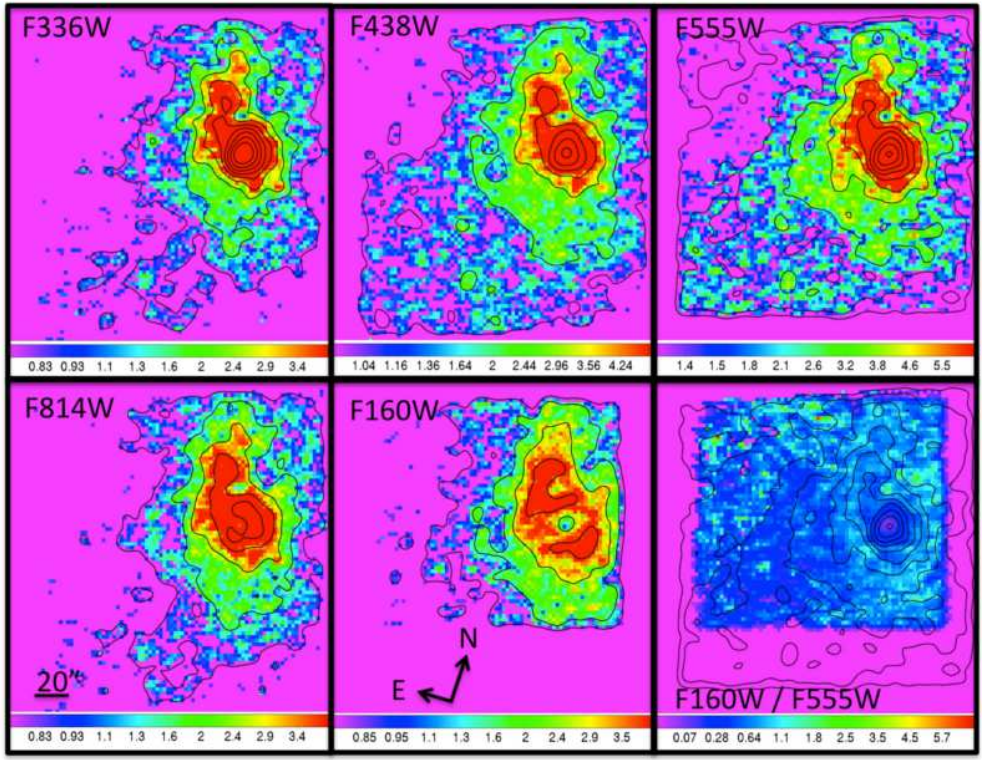}
\caption{\label{f:dens} Stellar density charts of the stars found in the core of NGC~2070 in the filters F336W , F438W, F555W, F814W and F160W. In each map the highest level is 5 times higher than the lower one. The lower-right plot shows that ratio between the F160W and the F555W maps. Isodensity contours from the F555W map are plotted for reference. In all the maps North is up, and East is to the left.}
\end{figure}

\begin{figure}
\epsscale{1.0}
\plotone{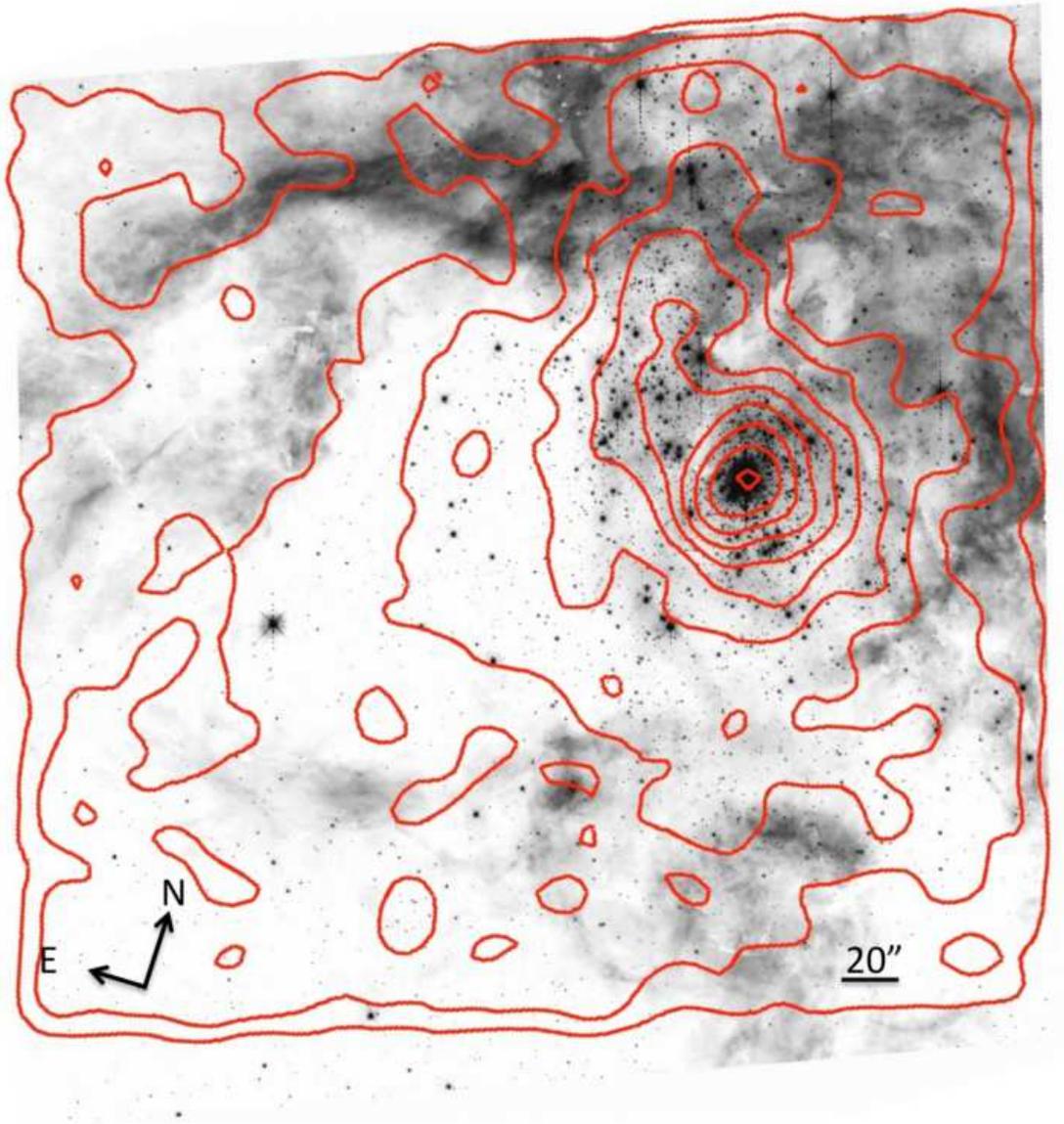}
\caption{\label{fig4nolan} UVIS image of 30~Dor in the F555W filter. North is up, East to the left. Isodensity contours as derived in Fig.~\ref{f:dens} are shown for reference.}
\end{figure}

Fig.~\ref{f:dens} shows the stellar density distributions in the  $\sim 2\arcmin 44\arcsec \times 2\arcmin 44\arcsec$ area around R136 in the filters F336W, F438W, F555W, F184W, and F160W. In each map the pixel scale is $2\arcsec$, and the density levels are the square root of the number of stars. The spatial distribution of the stars in the F336W, F438W, and F555W maps is remarkably similar, with the majority of the stars ``clustered'' at ${\rm R.A.}=05^{\rm h} 38^{\rm m} 42.6^{\rm s}, {\rm Dec}=-69\degr 06\arcmin 03\arcsec, {\rm J2000}$, which coincides with the center of 30~Dor (Figs.~\ref{fig4nolan} and~\ref{f:zoom}). The apparent decrease of star counts in the F814W and F160W maps where R136 is located is caused by the high crowding, since no completeness corrections were applied to the maps. 

In each map the density contours are not spherical. On the contrary they  show an arc that departs from R136 to the Northeast. This feature is clearly visible in all the filters, from the UV to the near IR, making it unlikely that local high reddening can be at the origin of this structure. The low impact of variable reddening in the stellar counts is also shown by the ratio between the F160W and F555W isodensity maps (Fig.~\ref{f:dens} -- lower-right panel).

In Fig.~\ref{f:cmds} the ($m_{\rm F555W}$, $m_{\rm F555W} - m_{\rm F814W}$) CMD of the stars found within $10\arcsec$ from the center of R136 is compared to that of the stars found in a  $14.4\arcsec \times  27.4\arcsec$ box centered on the Northeast clump (see also Fig.~\ref{f:zoom})s. In both plots, most of the brighter stars ($m_{\rm F555W} \la 19.5$) trace a well defined blue ($m_{\rm F555W} - m_{\rm F814W} \la0.5$) upper-MS, while below this value the majority of the stars show the red colors typical of intermediate and low-mass ($0.5\la M\la 5.3\, M_\odot$) PMS stars. Although the Northeast Clump is less then  $\sim 22\arcsec$ (corresponding to a projected distance of $\sim 5.4\, {\rm pc}$) away from R136, the redder and broader upper-MS of R136 indicates that the two populations are affected by different dust extinction. Isochrones fitting of the upper MS indicate that A$_{\rm F555W}=1.207$ is sufficient for the Northeast Clump, but for R136 A$_{\rm F555W}=1.525$ is more appropriate. The same values have been obtained by applying a procedure similar to that used in \citet{deMarchi2011}

It is commonly known that the steepness of the upper-MS and the paucity of massive stars makes it challenging to identify the MS turnoff (TO), and even harder to distinguish age differences of the order of few Myr, in star clusters younger than $\la 5\, {\rm Myr}$ using isochrone-fitting only. In such clusters a more robust age indicator is the luminosity of the Turn-On (TOn), defined as the  locus of the CMD where the PMS stars join the MS \citep{Stauffer1980, Belikov1998, Baume2003, Cignoni2010}. The age of a cluster can be defined as the time spent in the PMS phase by its most massive PMS star. 

The comparison between the CMDs of the two components shows that PMS stars in R136 are more than  $\sim 0.5$ mag brighter compared to those found in the Northeast Clump. Since R136 is affected by a higher reddening than the Northeast clump, we conclude that PMS stars in R136 are intrinsically brighter, and therefore that in R136 SF lasted for at least 1 Myr more than in the Northeast clump\footnote{We can exclude that the difference in magnitude between the two clumps is due to the distance, because $\Delta m_{\rm F555W}\simeq 1$ would imply that the Northeast clump would be at $\sim 79.5\, {\rm kpc}$, far beyond even the Small Magellanic Cloud.}. It is possible to see the Northeast clump also in Fig.~11 of  \citet{deMarchi2011}, where the spatial distribution of PMS stars younger than 4 Myr is compared to that of older objects with active mass accretion.

\section{Discussion and Conclusions}

We used deep HST/WFC3 UV, optical and near-IR observations of the core of 30~Dor to study its stellar content. In all the filters we identified a dual structure that cannot be attributed to dust extinction. A careful analysis of the stellar content of the clumps indicates that, while the Northeast clump formed the majority of its stars between 2 and 5 Myr ago, R136 started to form stars likely $\sim 2\, {\rm Myr}$ ago and was still active $\sim 1\, {\rm Myr}$ ago.

In retrospect, there are substantial indications in the literature that the Northeast Clump comprises a distinct, more evolved entity.  Its two brightest members, R141 
and R142, are early B supergiants that must be older than the denizens of R136 itself \citep{Feast1960, Parker1993}. Many of the fainter late-O/early-B stars, being evolved,  are also subject to the same age argument \citep{Walborn1997, Massey1998, Bosch1999}. In an analysis of star formation in the 30~Dor core based upon these observations, \citet{Selman1999} proposed three sequential events with mean ages less than 1.5~Myr, 2.5~Myr, and 5~Myr; their Figure~9 shows that the first is strongly concentrated toward R136, while the heaviest concentration of the intermediate and old objects coincides with the Northeast Clump.  

\begin{figure}
\epsscale{1.0}
\plotone{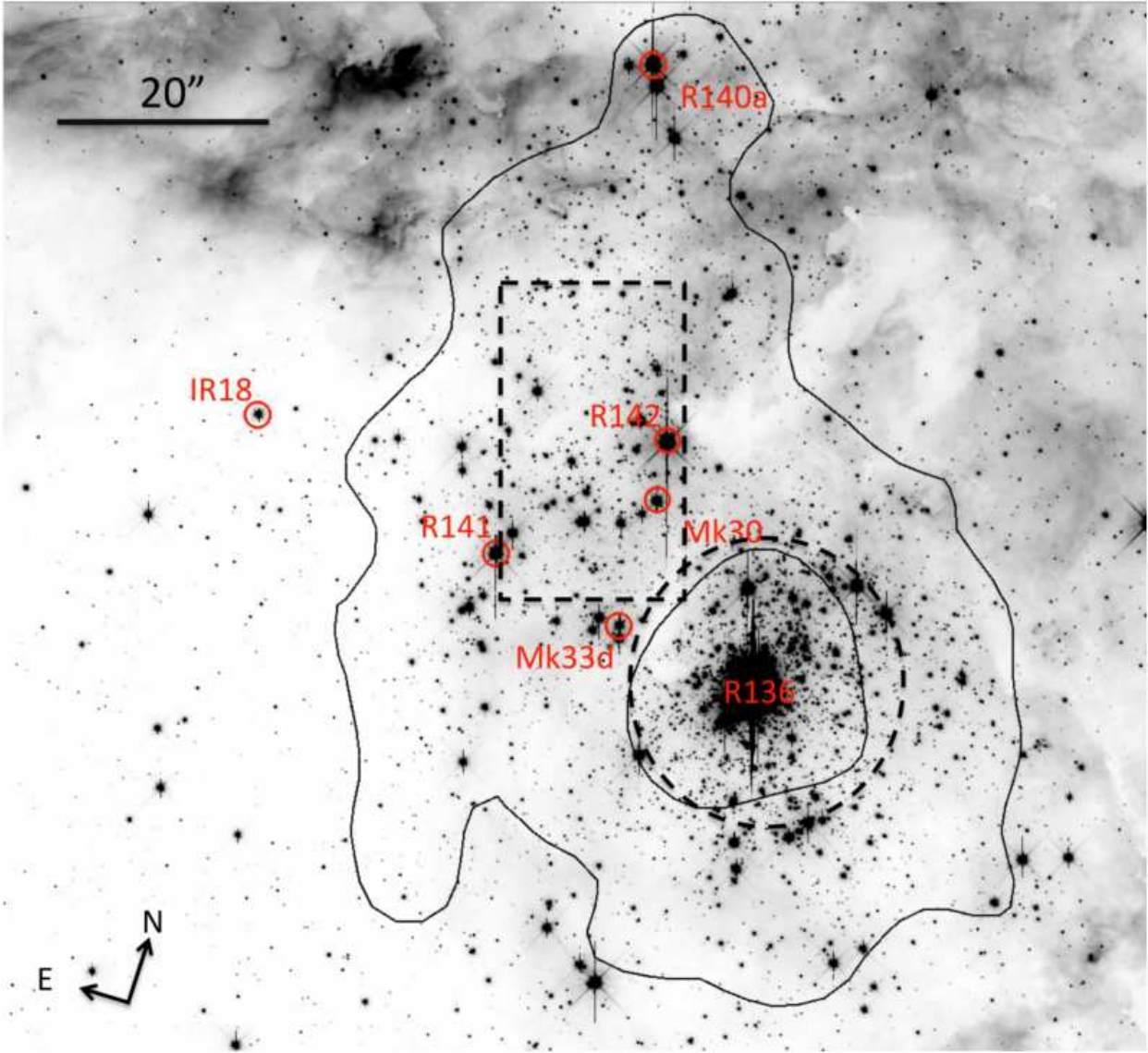}
\caption{\label{f:zoom} F555W UVIS image zoomed on the center of 30~Dor. The dashed circle and rectangle indicate the regions used to derive the CMDs shown in Fig.~\ref{f:cmds}. Stars mentioned in the Discussion are highlighted. Two isodensity contours are shown to guide the eye. 
}
\end{figure}

\begin{figure}
\epsscale{1.0}
\plotone{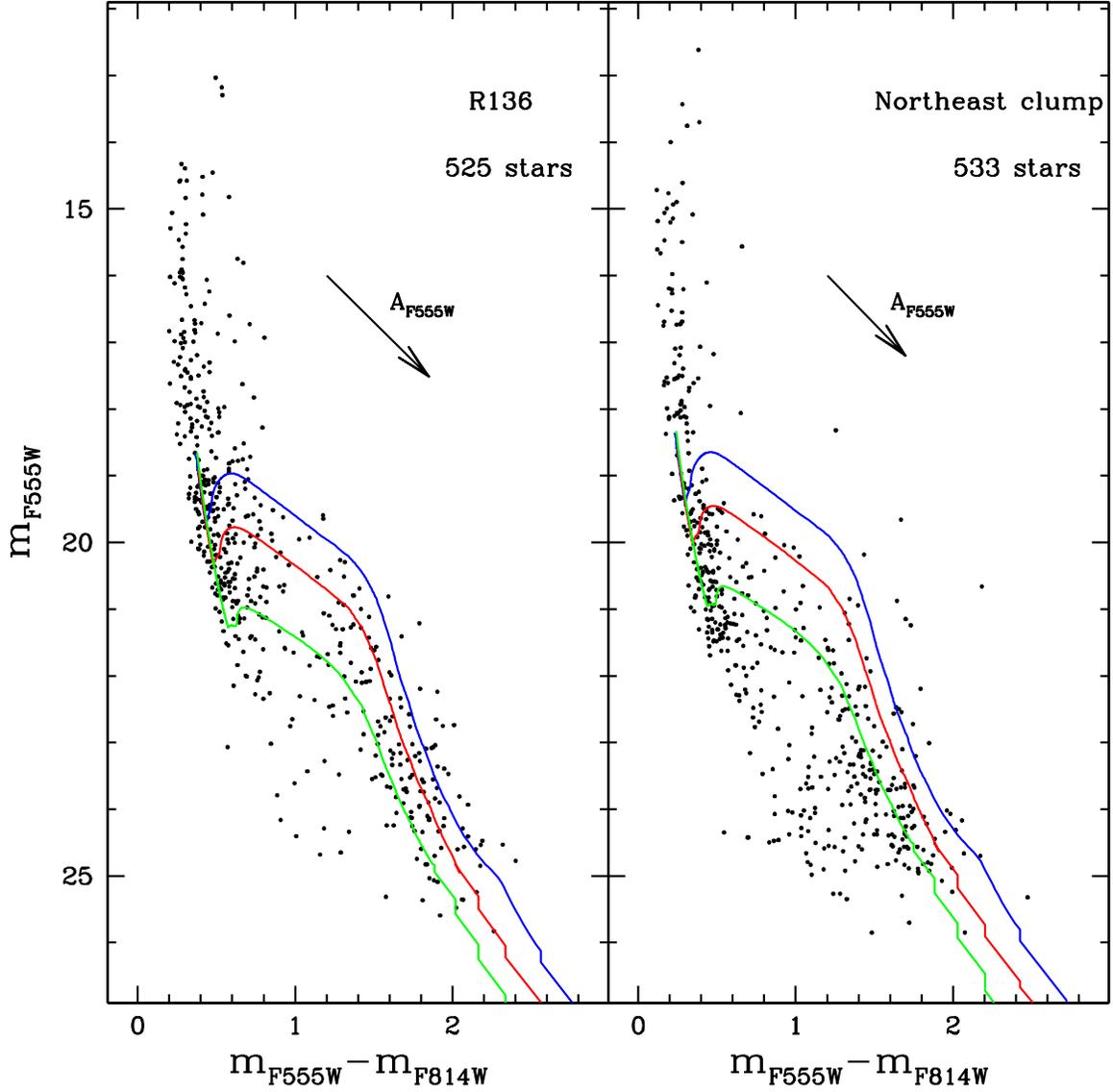}
\caption{\label{f:cmds} CMDs of $m_{\rm F555W}$ vs. $m_{\rm F555W} - m_{\rm F814W}$ of the stars with photometric error $<0.15$ mag in both the F555W and F814W filters found within $\sim10\arcsec$ from the center of R136 (left panel) and in a $\sim 14.4\arcsec \times 27\arcsec$ box centered on the Northeast clump (right panel).  Pisa isochrones for PMS stars with metallicity Z=0.008 and an age of 1, 2 and 5 Myr are plotted in blue, red and green respectively. In both the CMDs a distance modulus of 18.5 was assumed. R136 is better reproduced assuming an A$_{\rm F555W}=1.525$. For the Northeast clump A$_{\rm F555W}=1.207$ is sufficient.  }
\end{figure}

The age difference between R136 and the Northeast Clump can also explain  the, until now, puzzling presence of two WC stars in 30~Dor, as these highly evolved objects are seldom found in giant H{\sc ii} regions.  One of them, Mk~33d, is located in an apparent multiple system at the inner edge of the Northeast clump \citep{Melnick1985}, while the northward tail on the Northeast clump encompasses the multiple system R140 containing the other WC along with two WN's \citep{Moffat1987}.  Finally, even the isolated M supergiant  IR18 \citep{McGregor1981} somewhat further to the NE, long considered a field interloper, could actually be associated with the older clump. Positions of the mentioned stars with respect to the two sub-clusters are shown in Fig.~\ref{f:zoom}.  

This is, of course,  a complex three-dimensional region, and some earlier spectral types also appear in the predominantly older fields, particularly in the Mk~33 system.  The O2~If*/WN5 object Mk~30 \citep{Crowther2011} that lies adjacent to R142 along the ridge of the Northeast clump could be associated with R136; but on the other hand, some
such objects are also found in older young clusters \citep[e.g., TS3 in NGC~2060;][]{Schild1992}, possibly as a result of binary evolution. In summary, the information presented here provides strong support for our completely independent recognition of a double system with an age offset between them in the core of 30~Dor.

Characterizing the complex structure of massive star-forming regions such as NGC~2070 is important to understand the early stages of cluster formation. The age difference found between R136 and the Northeast clump, as well as their morphology, may indicate that the core of NGC~2070 is the result of a recent or ongoing merger between two sub-clusters. Interestingly from the analysis of the radial velocities of apparently single O stars, \citet[][submitted]{Brunet2012} found evidence for an internal rotation in NGC~2070, that can be due to either a recent merger between the main core of NGC~2070 and a secondary cluster or by the agglomeration and clumpiness of the gas. 

The majority of the stars studied in \citet{Brunet2012} coincides with R136, but 5 sources coincide with the Northeast Clump.  If the total mass of the system is $\sim 10^5\, {\rm M_\odot}$ \citep{Andersen2009} then the velocity difference, assuming a distance $d=4.5\, {\rm pc}$, would be $v=\sqrt(2GM/d)\simeq 13\, {\rm Km s^{-1}}$ for a hyperbolic encounter (i.e. 0 energy), and $\simeq 9\, {\rm km s^{-1}}$ for a circular orbit. The maximum velocity difference reported by \citet{Brunet2012} is $\sim 5\,  {\rm km s^{-1}}$. If R136 and the Northeast Clump are undergoing a merger, then either we are observing the interaction  with a low inclination (which would agree with the fact that we see the tidal tail on the plane of the sky), or the 2 systems are bound and are falling in with low velocity. A detailed analysis of the mass distribution in the two sub-clusters and a more sophisticated modeling of the two component interaction will be presented in two forthcoming papers (Sabbi et al. in prep, Gieles et al. in prep.)
 
The presence of distinct components in a very young system is also consistent with recent studies of galactic giant molecular clouds (GMC) \citep{Heyer2009}, which contradict the traditional paradigm that GMCs are gravitationally bound systems. \citet{Ballesteros2011} suggested that GMCs can undergo a hierarchical gravitational collapse, with the collapse occurring on scales ranging from individual cores to the whole cloud. In this scenario, because of the turbulent fragmentation of a GMC, SF will not be spread over the whole GMC, but will be localized in gravitationally bound pockets of gas \citep{Clark2005, Clark2008}. Star clusters will be the final products of the hierarchical merger of such smaller sub-structures \citep{Bonnell2003, Bate2009, Federrath2010}. 

We finally note that in a recent paper,  \citet{Fujii2012} suggested that the cuspy density profile of R136 and the large number of massive runaway stars that seem to escape from this cluster are signatures of a post-core-collapse star cluster \citep{Mackey2003}. Considering the small age of R136, such an early core collapse would occur only  if the cluster had an initial density of $\rho_c\ge 10^6 {\rm M_\odot}\, {\rm pc^{-3}}$, which is considerably higher than the current estimates.  \citet{Fujii2012}  suggest that an efficient way to speed up the collapse process is to invoke the hierarchical merging of several smaller clusters.

Observational evidence for a hierarchical process of SF comes also from the complex structure of massive star-forming regions such as NGC~604 in M33 \citep{jesus}, or NGC~346 in the Small Magellanic Cloud \citep{sabbi08}. Also, \citet{Gennaro2011} suggested that the elongation and mass segregation found in Westerlund~1 may be the footprint of a merger of multiple sub-clusters, formed almost coevally in the parental GMC. If massive star clusters form from mergers of smaller sub-systems, this can also explain the high fraction of rotating globular clusters found in our Galaxy. A study of the radial velocities and/or proper motions of the stars in the two clumps would provide a further test of the merging scenario. Such datasets are currently being obtained by us (e.g. H\'enault-Brunet et al. submitted; Lennon et al. in prep.).
 
%% In a manner similar to \objectname authors can provide links to dataset
%% hosted at participating data centers via the \dataset{} command.  The
%% second curly bracket argument is printed in the text while the first
%% parentheses argument serves as the valid data set identifier.  Large
%% lists of data set are best provided in a table (see Table 3 for an example).
%% Valid data set identifiers should be obtained from the data center that
%% is currently hosting the data.

%% In this section, we use  the \subsection command to set off
%% a subsection.  \footnote is used to insert a footnote to the text.


%% This section also includes several of the displayed math environments
%% mentioned in the Author Guide.


%% Included in this acknowledgments section are examples of the
%% AASTeX hypertext markup commands. Use \url without the optional [HREF]
%% argument when you want to print the url directly in the text. Otherwise,
%% use either \url or \anchor, with the HREF as the first argument and the
%% text to be printed in the second.

\acknowledgments

ES and DL are thankful to Chris Evans and Vincent H\'enault-Brunet for all the comments and suggestions that helped improve the letter. ES is also grateful to Aida Wofford and the Massive Stars and Starbursts Journal Club at STScI for the useful discussion and suggestions. The authors acknowledge support from grant GO-12499.01-A, and from NASA Hubble Fellowship grant HST-HF-51270.01-A to SdM, awarded by STScI, operated by AURA for NASA, contract NAS 5-26555. N.P. acknowledges partial support by  STScI-DDRF grant D0001.82435.

\clearpage

\end{document}